\documentstyle[12pt,aasms4,epsfig]{article}

\begin{document}

\begin{flushright}
PSU-PAL-99-1
\end{flushright}

\title{Cosmic-Ray Positrons: Are There Primary Sources?}

\begin{center} 
St\'ephane Coutu,$^\ast$\footnote{Corresponding author: coutu@phys.psu.edu}
Steven W.~Barwick,$^\dagger$
James J.~Beatty,$^\ast$
Amit Bhattacharyya,$^\ddagger$
Chuck R.~Bower,$^\ddagger$
Christopher J.~Chaput,$^\S$\footnote{Present address: Stanford Linear 
Accelerator, Stanford, California 94309, USA}
Georgia A.~de Nolfo,$^\ast$\footnote{Present address: Department of 
Physics, Downs Laboratory, California Institute of Technology, 
Pasadena, California 91125, USA} 
Michael A.~DuVernois,$^\ast$
Allan Labrador,$^\#$
Shawn P.~McKee,$^\S$
Dietrich M\"uller,$^\#$
James A.~Musser,$^\ddagger$
Scott L.~Nutter,$^\star$
Eric Schneider,$^\dagger$
Simon P.~Swordy,$^\#$
Gregory Tarl\'e,$^\S$
Andrew D.~Tomasch$^\S$ \&
Eric Torbet$^\#$\footnote{Present address: School of Natural Sciences, 
Institute for Advanced Study, Princeton, New Jersey 08540, USA}

\small

$^\ast${\it Departments of Physics and of Astronomy and Astrophysics, 
Davey Laboratory, Pennsylvania State University, University Park, 
Pennsylvania 16802, USA}

$^\dagger${\it Department of Physics, University of California at Irvine,
Irvine, California 92717, USA}

$^\ddagger${\it Department of Physics, Swain Hall West, Indiana University,
Bloomington, Indiana 47405, USA}

$^\S${\it Department of Physics, Randall Laboratory,
University of Michigan, Ann Arbor, Michigan 48109-1120, USA}

$^\#${\it Enrico Fermi Institute and Department of Physics, 
University of Chicago, Chicago, Illinois 60637, USA}

$^\star${\it Department of Physical Sciences, Eastern New Mexico University, 
Portales, New Mexico 88130, USA}

\end{center}

\normalsize

\noindent {\bf Abstract}

Galactic cosmic rays consist of primary and secondary particles.
Primary cosmic rays are thought to be energized by first order Fermi 
acceleration processes at supernova shock fronts within our Galaxy. The cosmic
rays that eventually reach the Earth from this source are mainly 
protons and atomic nuclei, but also include electrons. Secondary cosmic rays 
are created in collisions of primary particles with the diffuse interstellar
gas. They are relatively rare but carry important information on the Galactic 
propagation of the primary particles. The secondary component includes a
small fraction of antimatter particles, positrons and antiprotons. 
In addition, positrons and antiprotons may also come from unusual sources and 
possibly provide insight into new physics.
For instance, the annihilation of heavy supersymmetric dark matter particles
within the Galactic halo could lead to positrons or antiprotons
with distinctive energy signatures. With the High-Energy Antimatter 
Telescope (HEAT) balloon-borne instrument, we have measured the
abundances of positrons and electrons at energies between 1 
and 50~GeV. The data suggest that indeed a small additional
antimatter component may be present that cannot be explained by a purely 
secondary production mechanism. Here we describe the signature of the effect 
and discuss its possible origin.

\noindent {\it PACS codes:} 95.30.Cq, 95.85.Ry, 96.40.Cd, 96.40.De

\noindent {\it Keywords:} cosmic ray sources, positrons, electrons, 
supersymmetric dark matter, giant molecular clouds

\newpage

\section{Introduction}

Over the last 30 years, a number of efforts have been aimed at the study
of cosmic-ray electrons and positrons with balloon-borne instruments.
At energies between about 1 and 10~GeV, early
measurements [\cite{buf:pos,fan:pos}] found that the ``positron fraction'' 
e$^+$/(e$^+$+e$^-$) is essentially in agreement with a prediction 
[\cite{pro:pos}] where all positrons are assumed to be of secondary origin, 
and propagate according to the prescription of the simple ``leaky-box'' 
model of the Galaxy. This is illustrated in Figure~1a, which shows a 
compilation of measurements 
[\cite{buf:pos,fan:pos,agr:pos}-\cite{bar:apjl}]
 of the positron fraction as a function of energy
between 0.05~GeV and 50~GeV. The leaky-box prediction is shown as a solid
curve. At energies around 10~GeV and above, as shown in Figure~1a,
several measurements [\cite{agr:pos,mul:pos,go1:pos}] reported a 
significant excess of positrons over 
the fraction expected from secondary sources. This spurred a number of 
interpretations, ranging from inefficient production of primary electrons at 
high energies [\cite{bou:pos,tan:pos}], to hypothetical new sources of 
positrons [\cite{aha:pos}-\cite{tyl:pos}]. 

The HEAT balloon-borne instrument was designed and optimized to improve the
accuracy with which cosmic-ray electrons and positrons at energies from about 
1 to 50~GeV can be detected. 
The instrument and its performance during two balloon flights in 1994 and 1995,
respectively, are described in detail elsewhere 
[\cite{bar:apjl,heat:apj}-\cite{bar:prl}]. A compilation of positron
fraction measurements is shown in Figure~1a, where the HEAT results from the 
two flights combined are shown as filled squares. The overall proton rejection
factor achieved was nearly 10$^5$. Backgrounds due to atmospheric secondary
electrons and positrons were estimated by Monte-Carlo techniques, and compared
with measured growth curves [\cite{heat:apj}]. Such backgrounds amounted to 
1-2\%, and 20-30\% of the total electron and positron signals, respectively.
The uncertainty in the secondary corrections translated in a 
systematic uncertainty of $\sim 0.01$ in the positron fraction, comparable to 
the statistical uncertainty; however, any systematic error in the correction
would affect all data similarly, resulting in an overall normalization shift 
in the positron fraction distribution, preserving any structure observed.

\section{Secondary Production}

The HEAT results shown in Figure~1a did not confirm the previously-reported 
rise in the positron fraction starting at about 10~GeV. However, the data 
deviate
from the predictions of a purely secondary production mechanism in two ways. 
First, at energies below about 5~GeV the positron fraction was in excess of 
the expectations. For this low-energy energy region, another recent 
measurement [\cite{bar:cap}] also reported a positron fraction that was 
significantly higher than measured in the 1960's and 1970's. A possible
explanation of this effect would come from a solar modulation mechanism
that depends on the charge sign of the particle and changes from one solar 
cycle to the next [\cite{cle:pos}].

The second feature of the HEAT results is an indication of some structure in 
the energy dependence of the positron fraction above 7~GeV. This cannot be 
easily explained in terms of conventional secondary production mechanisms. As 
shown in Figure~1b, a slight enhancement in the positron fraction 
between about 7 and 20~GeV is observed, which may suggest a primary source of 
high-energy positrons. This feature appears in the
HEAT data from each flight taken separately. Figure~1b shows two predictions 
for interstellar secondary production in the energy region of interest. First,
the leaky-box prediction [\cite{pro:pos}] is shown as a solid red curve. A band
of uncertainty in this prediction due to the various uncertainties on the 
parameters of the model and that of the overall normalization is indicated
(hatched area). In this model, the spectrum of cosmic-ray positrons from 
secondary sources is calculated in the leaky-box approximation from:

\begin{equation}
j_{e^+}(E)={n c \over 4 \pi} \left( {dE \over dt} \right)^{-1}
\int_{E}^{\infty} dE' P_e(E') \times \exp \left[ -\int_{E}^{E'}
{dE'' \over \langle t(E'') \rangle (dE/dt)} \right] ,
\end{equation}

\noindent where $\langle t(E) \rangle$ is the mean cosmic-ray age at energy 
$E$, related to the rigidity-dependent mean Galactic escape length, 
$n$ is the mean density of interstellar nuclei, $P_e(E)$ is the rate of 
production of positrons in interstellar nuclear interactions, and $(dE/dt)$ 
is the rate of energy loss from synchrotron, 
inverse Compton, bremsstrahlung, and ionization processes. The positron 
fraction is obtained by dividing the predicted positron spectrum by the 
measured all-electron spectrum. A more recent calculation [\cite{str:pos}], 
shown as a dashed curve in Figure~1b, uses a more realistic Galactic 
diffusion model to predict the positron fraction from secondary production.
Qualitatively, it predicts the same behavior, a smooth, monotonic decrease of
the positron fraction without spectral features.
The HEAT data cannot be well fit by the secondary-production curves of 
Figure~1b. The confidence level
for the leaky-box prediction is essentially zero ($\chi^2$=96.5 for 9 degrees
of freedom), while that for the diffusion prediction is 0.9\% (after 
adjustments to take into account statistical runs in the data 
[\cite{roe:pro}]).
Although the band of uncertainty in the predictions is wide, all smooth
curves within it yield a similarly poor agreement with the data. 
If the structure seen in the data is real, it would indicate the onset of 
something new, such as an exotic source of high-energy positrons.
Here we consider several possible models.

\section{Sources of Primary Cosmic-Ray Positrons}

\subsection{Annihilating Dark Matter WIMPs}

First, it has been proposed that annihilating 
Galactic-halo dark-matter WIMPs (Weakly-Interacting Massive Particles) are a 
source of high-energy positrons 
[\cite{tur:ind,tur:dir,tyl:pos,jun:sdm,bal:ind}]. 
As most dark matter candidates are Majorana particles, direct annihilation
into e$^+$e$^-$ pairs is suppressed. In order to account for an observable
e$^+$e$^-$ line, a large total WIMP annihilation cross-section is required.
The WIMP density would then likely be low and not a major contributor to the
present-day cosmological mass density [\cite{tyl:pos}]. One
exception is a model by Kamionkowski and Turner (hereafter referred to as
KT) [\cite{tur:ind}] in which WIMPs with mass
m$_{\tilde{\chi}}$ greater than 80~GeV/c$^2$
or 91~GeV/c$^2$ can annihilate through resonant production of W$^+$W$^-$ or
Z$^0$Z$^0$ pairs. The resulting electrons and positrons are propagated in
a leaky-box model.
The model predicts enhancements in the positron fraction 
near energies of m$_{\tilde{\chi}}$/2 (due to direct decays of the gauge
bosons into e$^\pm$), and m$_{\tilde{\chi}}$/20 (continuum radiation due to 
more complex decay chains through intermediate production of $\tau ^\pm$, 
$\pi ^\pm$, quarks, etc.). If the experimental feature we observe is real, it 
could be a signature for the low-energy continuum radiation peak at around
m$_{\tilde{\chi}}$/20. Figure~2a shows a comparison of the positron fraction
we have measured with the HEAT instrument and a model prediction including
a WIMP-annihilation contribution. On the figure, the dashed curve is a
baseline secondary production distribution. To this baseline,
we add a contribution from annihilating dark matter neutralinos 
[\cite{tur:ind}]
of various masses, with an amplitude factor left as a fit parameter. This
amplitude factor is highly uncertain in the model, owing to a combination of
large uncertainties in the WIMP annihilation details and the astrophysical
parameters. The best-fit curve is shown as a solid line in Figure~2a, and 
occurs for a
neutralino mass of 380~GeV/c$^2$. The best-fit WIMP source strength is 1.8 
times greater than the estimated amplitude of the effect in the model,
well within the uncertainties in the prediction. The fit results are
summarized in Table~1. The resulting confidence level of 74\% is
markedly better than for fits to strictly secondary production models.

In recent work by Baltz and Edsj\"o (hereafter referred to as BE)
[\cite{bal:ind}], positron production
by annihilating dark matter neutralinos is revisited, and a large fraction
of the Minimal Supersymmetric Standard Model (MSSM) parameter space is
sampled. Again, decays and/or hadronization of the annihilation products
are simulated, and positron fluxes calculated, but a more complex diffusion 
model than in the KT scenario is used for the propagation of the
electrons and positrons. Here again, the predicted enhancement in the
positron fraction is allowed to be renormalized by a factor that is obtained
by fitting to the HEAT data. Two typical resulting best fit curves are shown
in Figure~2a, as dotted and dot-dash curves for 336 or 130~GeV/c$^2$ 
neutralinos,
respectively (for details of assumed MSSM parameters for these and other
models, see [\cite{bal:ind}]), and the fit results are summarized in Table~1.
Once again, an improvement is obtained compared to secondary models, but the
resulting confidence levels of 22\% and 42\% are not as high as the best-fit
KT model; this is mainly a result of the different propagation model used.

\subsection{Pair Creation Near Discrete Sources}

Second, primary positrons could arise when e$^+$e$^-$ pairs 
are created by electromagnetic processes, for instance through the conversion 
of high-energy $\gamma$ rays in the 
polar cap region of Galactic radio pulsars [\cite{har:pos}]. In this model, the
positron production rate $P_e(E)$ of equation~(1) is replaced by:
\begin{equation}
P^{HR}_{e}(E)=(1+kE^{0.5})P_e(E),
\end{equation}
where $k$ is given in terms of the Galactic pulsar birth rate $b_{30}$
(in units of 30~yr), the effective time $t_{max}$ during which the pulsar
emits $\gamma$ rays (in units of $10^4$~yr), the ratio $f_+$ of positrons
escaping the pulsar per $\gamma$ ray produced, and the total interstellar
mass $M$ (in units of $5 \times 10^9$ solar masses), by:
\begin{equation}
k=0.37 {b_{30}f_+t_{max}^{0.15} \over M}.
\end{equation}
\noindent By enhancing the baseline positron fraction from  
secondary sources with this kind of contribution, with $k$ left as a fit
parameter, we obtain the best-fit curve shown as a dashed line in Figure~2b, 
with $k=0.15$, comparable to reasonable expectations [\cite{har:pos}]. The 
fit results
are summarized in Table~1. Although the resulting confidence level of 50\%
is larger than that for purely secondary sources, 
the shape of the enhancement, a slow monotonic rise with energy, is not as 
compatible with the data as the local-enhancement effect obtained with some
WIMP-annihilation models. This model predicts that the positron fraction
should rise with energy beyond 10~GeV, reaching an asymptotic value of
0.5. This could be verified with measurements extending to higher energies.

Another electromagnetic process would be the interaction of very high-energy
$\gamma$ rays with optical and/or UV radiation in the vicinity
of discrete sources [\cite{aha:pos}],
resulting in e$^\pm$ pair production. In this scenario, the positron
production rate $P_e(E)$ of equation~(1) becomes:
\begin{equation}
P^{AA}_{e}(E)=P_e(E)+581.8 \times {\tau _{\gamma \gamma} \over E_{th}^{2.1}}
{\exp (-1/(x-1)) \over x(1+0.07x^{2.1}/\ln x)},
\end{equation}
\noindent
where $E_{th}=(m_e c^2)^2 /\epsilon _0$ is the threshold energy for gamma rays
interacting with ambient photons with characteristic energy $\epsilon _0$,
$x=4E/E_{th}$, $\tau _{\gamma \gamma}$ is the optical depth accumulated by the
gamma ray before escaping the source (a free parameter), and the numerical 
factor is calculated from formulas assuming a gamma-ray power-law index 
$\alpha \sim 2.1$. By adding a primary positron component from this effect
to the baseline from secondary sources, assuming various mean values for
the parameter $\epsilon_0$ and allowing the strength of the source to remain
a free parameter, we obtain the best-fit curve shown as a dotted line in 
Figure~2b.
The best fit occurs for $\epsilon _0 = 30$~eV, as summarized in Table~1,
which is in agreement with reasonable expectations [\cite{aha:pos}], and 
requires a
relatively weak source strength. The resulting confidence level of 75\% is
once again better than that for purely secondary sources.

\subsection{Positron Production in Giant Molecular Clouds}

A third possibility is the generation of electrons and positrons in hadronic
processes. In one model [\cite{dog:pos}], hadronic cosmic rays can enter and 
interact within giant molecular gas clouds, 
resulting in the secondary generation of mostly $\pi^\pm$ and $K^\pm$,
which ultimately decay into muons, and thereafter into electrons
and positrons. Fermi reacceleration due to fluctuations in the magnetic field 
in the turbulent gas could then boost the energy of the e$^\pm$.
In this model, if the typical field
strength in the cloud is $B$ and the minimum turbulence scale is $L_{min}$,
a characteristic magnetization momentum $p^* = eBc/L_{min}$ is defined.
Particles with momentum greater than $p^*$ tend to escape the cloud,
so that the spectrum of particles accelerated inside the cloud shows an
enhancement near $p^*$. For reasonable choices for the parameters in the
model, it is possible to obtain $p^* = 10$~GeV/c, and a positron
fraction curve is obtained with an enhancement starting near 10~GeV.
If we add to the baseline secondary positron fraction such a primary
component from giant molecular clouds,
and allow the strength of the effect to be a free parameter, the resulting
best fit is the solid curve of Figure~2b. A relatively weak source
is sufficient (see Table~1) to fit the data with a confidence level of 80\%.

\subsection{Other Positron Sources}

Other primary positron sources have been suggested as well. For example,
e$^+$e$^-$ pair production in the magnetosphere of pulsars could be followed 
by particle acceleration to relativistic energies in the pulsar wind driven 
by low-frequency electromagnetic waves [\cite{chi:pos}]. Or else $\beta^+$ 
radionuclei 
such as $^{56}$Co ejected during a supernova blast, possibly followed by shock 
acceleration in the envelope [\cite{ski:pos}], could result in an enhanced 
high-energy positron population. The uncertainties in the models and in the 
data are such that none of these models can yet be ruled out.

\section{Conclusions}

In conclusion, the HEAT positron measurements indicate a subtle feature,
which cannot be caused by the atmospheric corrections applied to the data
or any other known systematic effect. If confirmed by future measurements, 
this feature could be evidence for
a new exotic source of positrons, especially at energies beyond about
7~GeV. The exact nature of this source cannot be determined until
higher-statistics measurements of the positron fraction near 10~GeV,
as well as extensions of the measurements to energies beyond 50~GeV,
become available. All models proposed and discussed here are unconventional if
not contrived. Only through detailed studies of the exact shape of the 
spectral features in the positron fraction can we expect to determine which,
if any, of the models offers the correct description.

\noindent
{\bf Acknowledgements}

We are grateful to M.~Kamionkowski for providing us with his computer code
for calculating positron-fraction enhancements from annihilating WIMPs.
This work was supported by NASA and by financial assistance from our 
universities.

\newpage

\begin{figure}
\centering
\epsfig{file=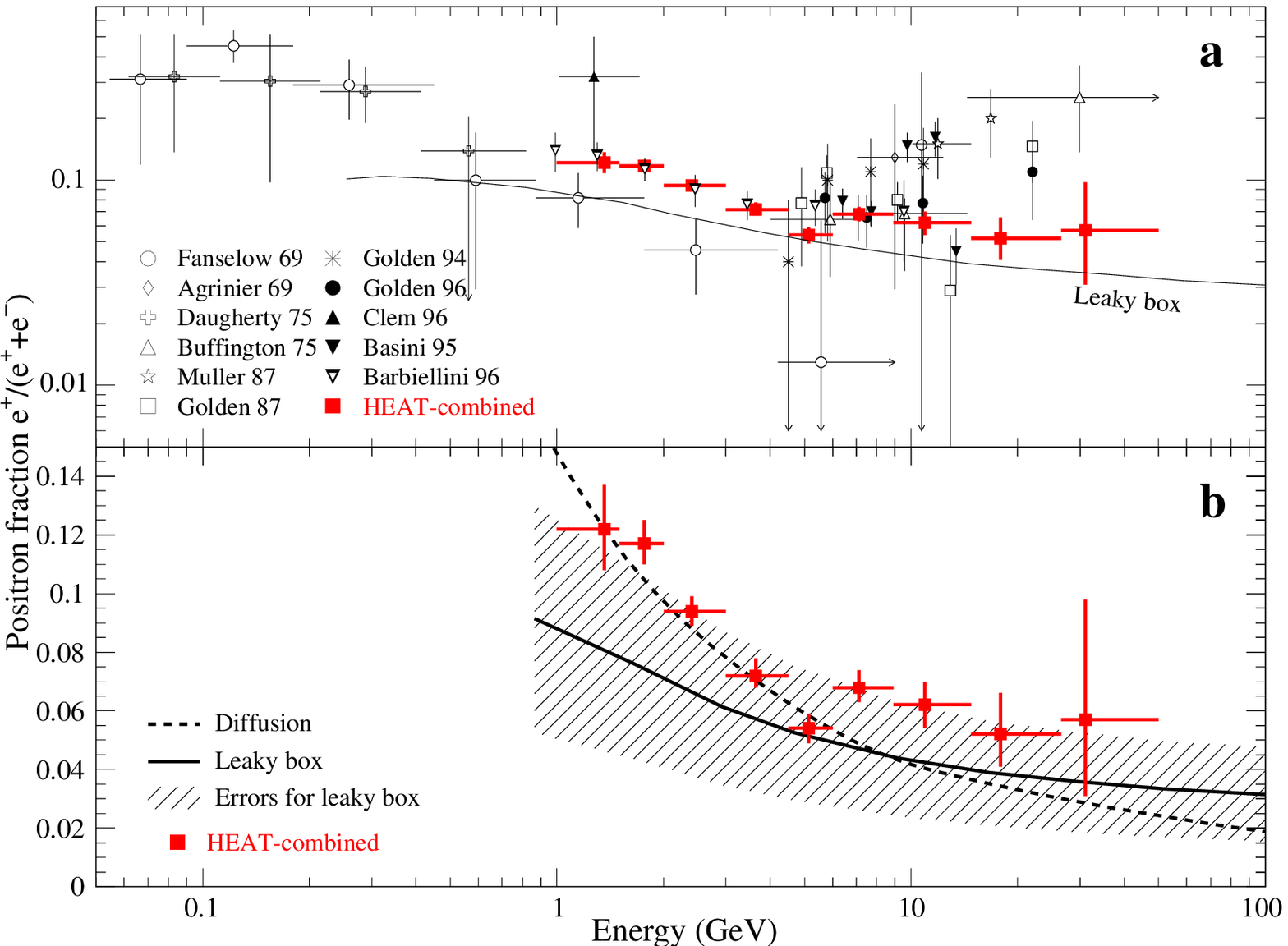,width=14cm}
\caption{{\bf a} Compilation of measurements [\cite{buf:pos,fan:pos,agr:pos}
-\cite{bar:apjl}] of the positron fraction between
0.05 and 50~GeV. The solid curve is a model calculation [\cite{pro:pos}]
assuming that
all positrons are from secondary sources, and propagate according to a simple
Galactic leaky-box model. ``HEAT-combined'' refers
to the combination [\cite{bar:apjl}] of the data sets from the two HEAT 
flights.
{\bf b} The positron fraction measured with the HEAT instrument, shown on
a vertical linear scale. The solid curve is a leaky-box secondary model
prediction [\cite{pro:pos}], surrounded by an estimated band of uncertainty 
shown as the cross-hatching. The dashed curve is a secondary model 
prediction using Galactic diffusion [\cite{str:pos}]. 
} \label{pfr_sec}
\end{figure}

\begin{figure}
\centering
\epsfig{file=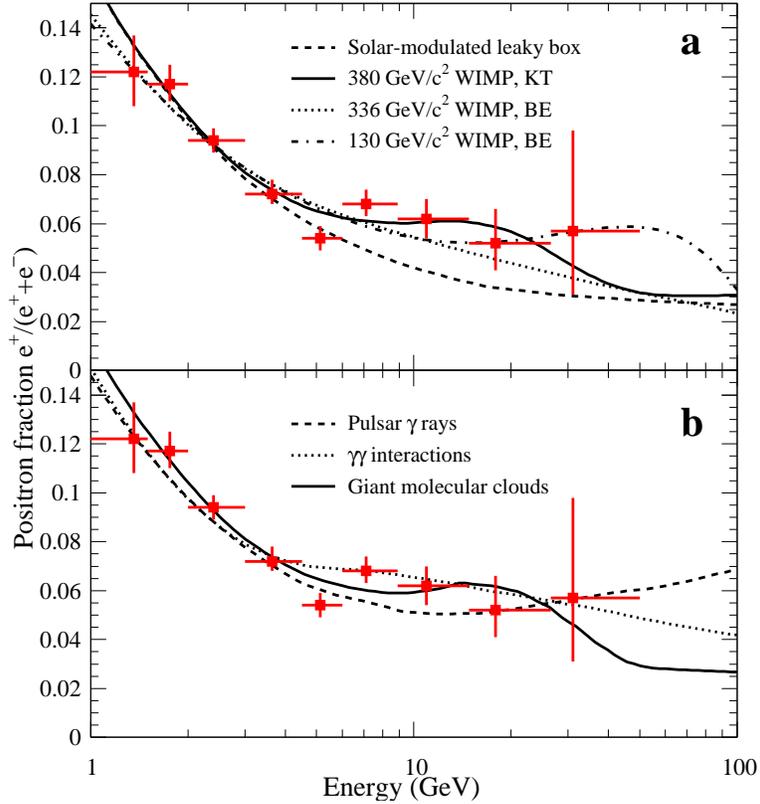,width=10cm}
\caption{{\bf a} The HEAT positron fraction compared with best-fit model 
predictions
with an additional positron component arising from annihilating dark matter
neutralinos. The dashed curve is the baseline solar-modulated leaky-box
secondary-production prediction [\cite{cle:pos}], renormalized by a factor of 
0.85. The solid curve shows an increased positron content due to annihilating
380~GeV/c$^2$ neutralinos in the model of Kamionkowski and Turner 
[\cite{tur:ind}]. 
The dotted and dot-dash curves show an increased positron content due to 
annihilating 336 or 130 GeV/c$^2$ neutralinos, respectively, in the model of 
Baltz and Edsj\"o [\cite{bal:ind}]. 
{\bf b} The HEAT positron fraction compared with best-fit model predictions 
from
astrophysical sources of positrons that are in addition to secondary production
mechanisms. The dashed curve is the positron enhancement resulting from
high-energy $\gamma$ rays converting to e$^+$e$^-$ pairs near the magnetic
poles of pulsars [\cite{har:pos}]. The dotted curve represents a positron 
enhancement due 
to high-energy $\gamma$ rays interacting with low-energy optical or 
UV photon fields [\cite{aha:pos}]. The solid curve shows the enhancement from
cosmic-ray interactions within giant molecular clouds [\cite{dog:pos}].
} \label{pfr_pri}
\end{figure}

\newpage

\begin{table}[h]
\vspace{-12pt}
\caption{Statistical agreement between HEAT positron fraction results
and various primary positron source models.\label{statpri}}
\begin{center}
\begin{tabular}{cccc}
\hline\hline
Model & Fit parameter & Source amplitude factor & Confidence level \\
\hline
KT WIMPs & m$_{\tilde{\chi}} = 380$~GeV/c$^2$ & $1.81\pm0.53$ & 74\% \\
BE WIMPs & m$_{\tilde{\chi}} = 336$~GeV/c$^2$ & 11.7 & 22\% \\
BE WIMPs & m$_{\tilde{\chi}} = 130$~GeV/c$^2$ & 54.6 & 42\% \\
Pulsar $\gamma$ rays & $k=0.15$ & & 50\% \\
$\gamma \gamma$ interactions & $\epsilon_0 = 30$~eV & $0.0262\pm0.0076$
& 75\% \\
Giant molecular clouds & & $0.097 \pm 0.029$ & 80\% \\
\hline
\hline
\end{tabular}
\end{center}
\end{table}

\noindent The ``source amplitude factor'' is an arbitrary normalization that 
indicates the best-fit strength of the effect compared to the one predicted 
by the authors of the model.

\end{document}